\documentstyle[twocolumn,prl,aps,epsfig,amssymb]{revtex}

\def\pslash{\rlap{\hspace{0.02cm}/}{p}}
\begin{document}
\draft
\title{The study of a flavor-changing neutral toppion production
process $e^+e^-\rightarrow t\bar{c}\Pi_t^0$}

\author{Xuelei Wang $^{1,2}$, Yueling Yang$^{2}$, Bingzhong Li$^2$,
Chongxing Yue $^{1,2}$, Jinyu Zhang$^2$}

\address{$^1$ CCAST (World  Laboratory), P.O.Box 8730, Beijing 100080, China}
\address{$^2$ College of Physics and Information Engineering, Henan Normal University, Henan 453002, China}
\date{\today}
\maketitle

\begin{abstract}
 We have studied a flavor-changing toppion production process
 $e^{+}e^{-}\rightarrow t\overline{c}\Pi^{0}_{t}$ in
 the topcolor-assisted technicolor(TC2) model. The studies show that,
 with high centre of mass energy in TESLA collider, the production cross
 section of $e^{+}e^{-}\rightarrow t\overline{c}\Pi^{0}_{t}$ is at the order
 of magnitude 0.1 fb in most parameter regions of TC2 model and a few tens
 events of toppion can be produced each year. The resonance
 effect can enhance the cross section to a few fb when toppion mass is small.
 With clean background, the toppion events can possibly be detected
 at TESLA collider. On the other hand, we find that there exists a narrow
 peak near $m_t-m_c$ in the toppion-charm invariant mass distribution which
 could be clearly detected. Therefore, such a toppion
  production process  $e^{+}e^{-}\rightarrow t\overline{c}\Pi^{0}_{t}$
  provides a unique chance to detect toppion events and test the TC2
  model.

\end{abstract}

\vspace{.5cm}
\noindent{\bf I. Introduction}

~~ A unified and beautiful
description of the weak and electromagnetic interaction is given
by the Glashow -Weinberg-Salam(GWS)
theory based on the gauge group $SU_{L}(2)\times U_{Y}(1)$. Many
experimental successes have been credited to this model. However,
its symmetry breaking sector is unclear. Probing the electroweak
symmetry breaking mechanism will be one of the most important
tasks at future high energy colliders.

Dynamical electroweak symmetry breaking, for example
technicolor(TC) type theories\cite{TC}, is an attractive idea that
avoids the shortcomings of triviality and unnaturalness  arising
from the elementary Higgs field of GWS theory. The simplest
QCD-like extended technicolor model\cite{QCD-like} leads to a too
large oblique correction S parameter\cite{S-parameter}, and is
already ruled out by the CERN $e^{+}e^{-}$ collider LEP precision
electroweak measument data \cite{LEP1,LEP2}. Various improvements
have been proposed to make the predictions consistent with the LEP
precision measurement data. A more realistic TC model is
topcolor-assisted technicolor(TC2) model \cite{TC2}, which can
also solve heavy top quark problem. In TC2 theory, the electroweak
symmetry breaking(ESB) is driven mainly by technicolor
interactions, the extended technicolor give contributions to all
ordinary quark and lepton masses including a very small portion of
the top quark mass: $m^{'}_{t}=\varepsilon m_{t}$ $(0.03\leq
\varepsilon \leq 0.1)$\cite{Burdman}. The topcolor interactions
also make small contributions to the ESB and give rise to the main
part of the top mass $(1-\varepsilon)m_{t}$ . One of the most
general predictions of TC2 model is the existence of three
Pseudo-Goldstone Boson in a few hundred GeV region, so called
toppions: $\Pi^{0}_{t}$,$\Pi^{\pm}_{t}$. The physical particle
toppions can be regarded as the typical feature of TC2 model.
Thus, studying the possible signature of toppions and toppion
contributions to some processes at the high energy colliders is a
good method to test TC2 model. There have been a lot of
literatures  related to this field
\cite{toppion,flavor-changing,He}. Another feature of TC2 model is
the existence of flavor-changing coupling of neutral toppion to
top and charm quarks.
 As it is know, there is no flavor-changing neutral current(FCNC)
at tree-level in the standard model(SM). The FCNC processes in SM can hardly
be detected. Any observation of the flavor-changing coupling deviated
from that in the SM would be unambiguously signal of new physics. So,
the study of some processes including $\Pi^{0}_{t}-t-c$ vertex
within the framework of TC2 model would provide some information
of the flavor-changing coupling and a feasible method to detect
the signals of toppion.

In this paper, we will study a neutral toppion production process
$e^{+}e^{-}\rightarrow t\overline{c}\Pi^{0}_{t}$ which includes
the flavor-changing vertex $\Pi^{0}_{t}-t-c$. Our results
show that, with favorable parameter values, the production cross
section of $e^{+}e^{-}\rightarrow t\overline{c}\Pi^{0}_{t}$
 is expected to reach the order of magnitude of 10 fb. The signals of
toppion might be
detected experimentally at DESY TESLA with high energy and
large luminosity.

This paper is organized as follows. In sec II, we shall present the
calculations of the production cross section of the process
$e^{+}e^{-}\rightarrow t\overline{c}\Pi^{0}_{t}$. The numerical
results of the cross section and the concluding remarks will be
presented in sec III.

\vspace{.5cm}
\noindent{\bf II The calculations of the production cross
section}

For TC2 models, the underlying interactions, topcolor
interactions are non-universal and therefore does not posses a GIM
mechanism. This is an essential feature of this kind
of models due to the need to single out the top quark for
condensation. The non-universal gauge interactions result in the
new flavor-changing coupling vertices when one writes the
interactions in the quark mass eigen-basis. Thus, the toppions
predicted by this kind of models have large Yukawa couplings to the
third generation and can induce the new flavor-changing couplings.
The relevant flavor-changing vertices including the large
top-charm transition for the neutral toppion can be written
as\cite{He}
\begin{eqnarray*}
i\frac{m_{t}\tan\beta}{v_{\omega}}K^{tc}_{UR}K^{tt^{*}}_{UL}
\overline{t_{L}}c_{R}\Pi^{0}_{t}+h_{.}c_{.}
\end{eqnarray*}
where $\tan\beta=\sqrt{(\frac{v_{\omega}}{v_{t}})^{2}-1}$,
$v_{\omega}=246$ GeV is electroweak symmetry-breaking scale and
$v_{t}\approx 60-100$ GeV\cite{He} is the toppion decay
constant. $K^{tt}_{UL}$is the matrix element of the unitary  matrix
$K_{UL}$ which the CKM matrix  can be derived as
$v=K^{-1}_{UL}K_{DL}$ and $K^{ij}_{UR}$ are the matrix elements
of the right-handed rotation matrix $K_{UR}$. Their values can be
written as:
 $$K^{tt^{*}}_{UL}=1 \hspace{2cm}K^{tt}_{UR}=1-\varepsilon$$
$$K^{tc}_{UR}\leq\sqrt{1-(K^{tt}_{UR})^{2}}\approx
\sqrt{2\varepsilon-\varepsilon^{2}}$$

With the flavor-changing coupling $\Pi^{0}_{t}-t-c$, the neutral
toppion can be produced associated with a single top quark at
$e^{+}e^{-}$ colliders, i.e., $e^{+}e^{-}\rightarrow t\overline{c}
\Pi^{0}_{t}$. The Feynman diagrams of the process is shown in
Fig.1.
\begin{figure}[htb]

\begin{center}
\epsfig{file=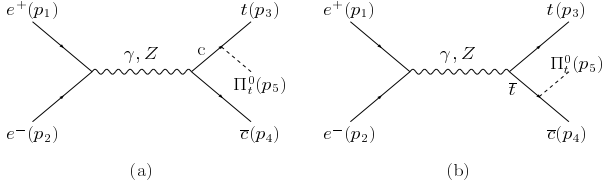,width=230pt,height=250pt}
\vspace*{-5cm}
 \caption{The Feynman diagrams of the process
      $e^{+}e^{-}\rightarrow t\overline{c}\Pi^{0}_{t}$.}
 \label{ee}
\end{center}
\end{figure}

The explicit expression of the production amplitudes is as
follows:
\begin{eqnarray*}
M&=&M^{z}_{a}+M^{z}_{b}+M^{\gamma}_{a}+M^{\gamma}_{b}\\
M^{z}_{a}&=&\frac{\sqrt{2}m_{t}\tan\beta}{2v_{\omega}}G_{F}M^{2}_{Z}
K^{tc}_{UR}K^{tt^{*}}_{UL}G(p_{3}+p_{5},m_{c})\\
& &G(p_{1}+p_{2},m_{z})
\overline{u}_t(p_{3})[-\frac{8}{3}s^2_{\omega}(\pslash
_{3}+\pslash_{5})\gamma^{\mu}R\\
& &+m_{c}(2-\frac{8}{3}s_{\omega}^{2}) \gamma^{\mu}L]
v_{c}(p_{4})\overline{v}_{e^{+}}(p_{1})\gamma_{\mu}\\ & &
(-L+2s_{\omega}^{2})u_{e^{-}}(p_{2})\\
M^{z}_{b}&=&\frac{\sqrt{2}m_{t}\tan\beta}{2v_{\omega}}G_{F}M^{2}_{Z}
K^{tc}_{UR}K^{tt^{*}}_{UL}G(p_{4}+p_{5},m_{t})\\
& &G(p_{1}+p_{2},m_{z})
\overline{u}_t(p_{3})\gamma^{\mu}[(2-\frac{8}{3}s_{\omega}^{2})L
(\pslash _{4}+\pslash_{5})\\
&  &-\frac{8}{3}s_{\omega}^{2}m_{t}R]
v_{c}(p_{4})\overline{v}_{e^{+}}(p_{1})\gamma_{\mu}(-L
+2s_{\omega}^{2})u_{e^{-}}(p_{2})
\\
M^{\gamma}_{a}&=&-\frac{8\sqrt{2}}{3}\frac{m_{t}\tan\beta}{v_{\omega}}
G_{F}M^{2}_{W}s_{\omega}^{2}
K^{tc}_{UR}K^{tt^{*}}_{UL}\\
& &G(p_{3}+p_{5},m_{c})G(p_{1}+p_{2},0) \overline{u}_t(p_{3})[(\pslash
_{3}+\pslash_{5})\gamma^{\mu}R\\
& &+m_{c}\gamma^{\mu}L]
v_{c}(p_{4})\overline{v}_{e^{+}}(p_{1})\gamma_{\mu}u_{e^{-}}(p_{2})\\
M^{\gamma}_{b}&=&-\frac{8\sqrt{2}}{3}\frac{m_{t}\tan\beta}{v_{\omega}}
G_{F}M^{2}_{W}s_{\omega}^{2}
K^{tc}_{UR}K^{tt^{*}}_{UL}\\
& &G(p_{4}+p_{5},m_{t})G(p_{1}+p_{2},0)
\overline{u}_t(p_{3})\gamma^{\mu}\\& &
(\pslash _{4}+\pslash_{5}
+m_{t})Rv_{c}(p_{4})\overline{v}_{e^{+}}(p_{1})\gamma_{\mu}u_{e^{-}}(p_{2})
\end{eqnarray*}
Here, $G(p,m)=\frac{1}{p^{2}-m^{2}}$ is the propagator of the particle, $L=\frac{1}{2}(1-\gamma_5)$,
$R=\frac{1}{2}(1+\gamma_5)$, $s^2_{w}=\sin^2 \theta_w $($\theta_w$ is the Weinberg angle).
 We can see
that the time-like momentum may hit the top-pole in the top quark
propagator. So we should take into account the effects of the
width of top quark in the calculations. i.e., we should take the
complex mass term $m_{t}^{2}-im_{t}\Gamma_{t}$ instead of the
simple top quark mass term $ m_{t}^{2}$ in the top quark
propagator. The $-im_{t}\Gamma_{t}$ term is important in the
vicinity of the resonance. As it is know, the top quark decays
almost to $W^{+}b$. So, the total top quark decay width can be
replaced approximately by $\Gamma(t\rightarrow W^{+}b)$ in our
calculation. With above production amplitudes, we can obtain the
production cross section directly.

\vspace{.5cm}
\noindent{\bf III. The numerical results and conclusions}

 In our calculations, we take $m_{t}=174$ GeV, $m_{c}=1.5$ GeV
$v_{t}=60$ GeV. There are two free parameters involved in the
production amplitudes, i.e., $\varepsilon,M_{\Pi}$. To see the
influence of toppion mass and parameter $\varepsilon$ on the
production cross section, we take the mass of toppion $M_{\Pi}$ to vary in certain
ranges 150 GeV$\leq M_{\Pi}\leq$ 350 GeV and
$\varepsilon=0.03,0.06,0.1$ respectively. One of the next
generation $e^{+}e^{-}$ colliders is TESLA in Europe\cite{TESLA}. In its first
stage, TESLA can run at centre of mass energy $\sqrt{s}=500$ GeV. An upgrade
to close to 1 TeV is planned. The predicted maximal luminosity
 can reach 500 fb$^{-1}$/year. In this paper, to study the
 discovery potential of $\Pi^{0}_{t}$ at TESLA, we will take
 $\sqrt{s}=500$ GeV, 800 GeV,1600 GeV
 respectively. The final numerical results of the cross section are
 summarized in Fig.2-4.

   The Fig.2-4 are the cross section plots as the function of $M_{\Pi}$
   for$\sqrt{s}=500,800,1600$ GeV and $\varepsilon=0.03,0.06,0.1$
   respectively. We can see that the cross section of $e^{+}e^{-}
   \rightarrow t\overline{c}\Pi^{0}_{t}$ increase with the mass of
   toppion $M_{\Pi}$ decreasing. Specially, when $M_{\Pi}<
   m_{t}-m_{c}$, the cross section increase sharply due
   to the resonance effect in Fig.\ref{ee}(b). The cross section can
   reach about a few fb for small $M_{\Pi}$.
  With the yearly integrated luminosity of $\pounds\sim
    500$ fb$^{-1}$ expected at TESLA, one could collect a few
    thousands of $\Pi_{t}$ events each year. With high energy($\sqrt{s}=800,1600$ GeV),
 the order of magnitude of the cross section is at 0.1fb in most parameter regions,
a few tens of $\Pi^{0}_{t}$
    events can be produced each year. As it is known, there is no
    flavor-changing coupling between charm and top quarks in SM. The
    background of $e^{+}e^{-}\rightarrow t\overline{c}\Pi^{0}_{t}$
    should be very clean. So, the events of neutral toppion
    produced in the process $e^{+}e^{-}\rightarrow t\overline{c}\Pi^{0}_{t}$
   should be detected at TESLA, specially, in case of light $\Pi^{0}_{t}$.
   Comparing the cross sections for
   $\sqrt{s}$=500,800,1600 GeV, we can conclude that $\sqrt{s}=500$
   GeV is a favorable centre of mass energy to detect light toppion
   events. But for heavy toppion, high energy can
   enhance the cross section. So, high energy is needed to detect
   the heavy toppion.

   Due to the resonance effect for light toppion, there should be
   a peak near $m_t-m_c$ in the toppion-charm invariant
   mass distribution. To show the above results, we plot, in
   Fig.5, the toppion-charm invariant mass distribution with
   $\sqrt{s}$=500 GeV, $M_{\Pi}=$150 GeV, $\varepsilon=0.1$. It
   can be seen that the resonance peak can possibly be observable.
   So, the study of the invariant mass distribution may provide
   another good way to detect signal of toppion.

     In conclusion, we have studied a flavor-changing process
      $e^{+}e^{-}\rightarrow t\overline{c}\Pi^{0}_{t}$ in TC2
      model. We find there are some features about this process.
      (1) For light toppion $(M_{\Pi}< m_{t}-m_{c})$, there exists the resonance
   effect which can enhance the cross section significantly. So, the
      process $e^{+}e^{-}\rightarrow t\overline{c}\Pi^{0}_{t}$ is
      favorable for light toppion detecting. High centre of mass is needed for
      heavy toppion detecting. With $\sqrt{s}$=800,1600 GeV, the cross
      section of $e^{+}e^{-}\rightarrow t\overline{c}\Pi^{0}_{t}$
      is at the order of magnitude of 0.1 fb in most parameter regions,
      a few tens $\Pi^{0}_{t}$
      events can be produced via $e^{+}e^{-}\rightarrow
      t\overline{c}\Pi^{0}_{t}$. (2) There is no
      flavor-changing coupling between charm-top quarks at tree
      level in SM, the background of this flavor-changing
      process $e^{+}e^{-}\rightarrow
      t\overline{c}\Pi^{0}_{t}$ might be very clean. So, there is
      reasonable hope to isolate the events experimentally with a
      few tens each year. Therefore, the $e^{+}e^{-}\rightarrow
      t\overline{c}\Pi^{0}_{t}$ process provides a feasible test
      of TC2 model with the detecting of toppion signals.

\vspace{.5cm}
\noindent{\bf Acknowledgments}

This work is supported by the National Natural Science Foundation
of China, the Excellent Youth Foundation of Henan Scientific
Committee, and the Henan Innovation Project for University
Prominent Research Talents.

\begin{figure}[htb]
\begin{center}
\epsfig{file=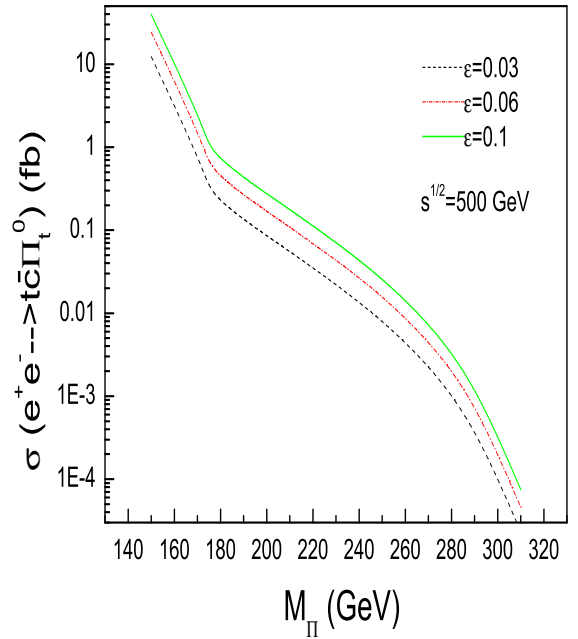,width=240pt,height=280pt} \caption{The
cross section of $ e^{+}e^{-}\rightarrow t\overline{c}
\Pi^{0}_{t}$ versus toppion mass $M_{\Pi^0_t}$(150-350 GeV) for
$\sqrt{s}=500$ GeV and $\varepsilon=0.03$(dash
line),$\varepsilon=0.06$ (short dash-dot
line),$\varepsilon=0.1$(solid line) repectively.} \label{fig2}
\end{center}
\end{figure}
\vspace*{-.2cm}

\begin{figure}[htb]
\begin{center}
\epsfig{file=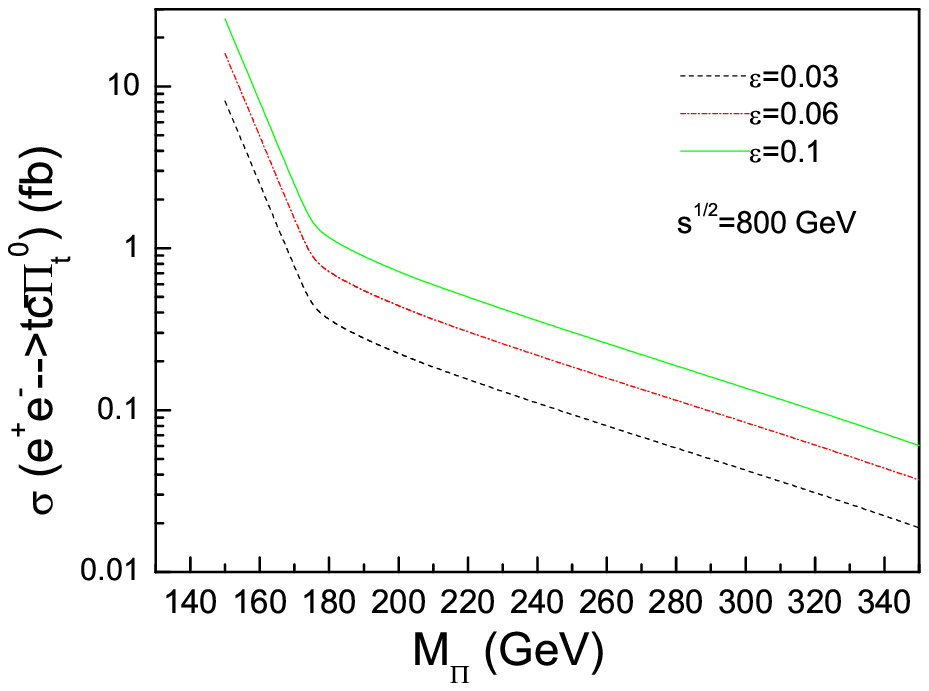,width=240pt,height=280pt} \caption{The same
plots as Fig.2 for $\sqrt{s}=800$ GeV with $\varepsilon=0.03$(dash
line),$\varepsilon=0.06$ (short dash-dot
line),$\varepsilon=0.1$(solid line) repectively  } \label{fig3}
\end{center}
\end{figure}
\vspace*{-.2cm}



\begin{figure}[htb]
\begin{center}
\epsfig{file=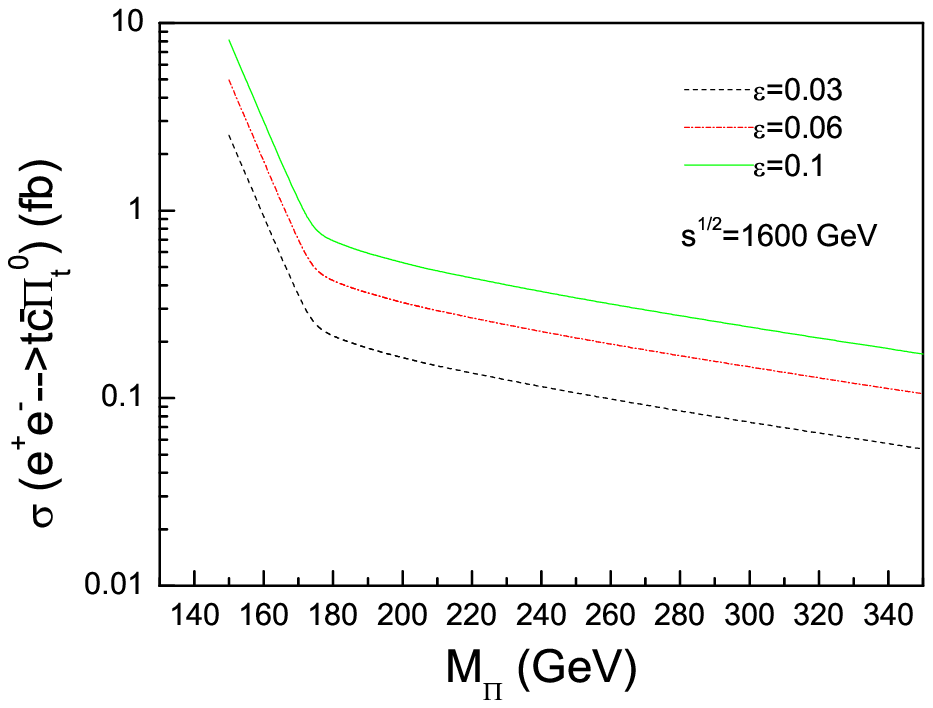,width=240pt,height=280pt} \caption{The
same plots as Fig.2 for $\sqrt{s}=1600$ GeV with $\varepsilon=0.03$(dash
line),$\varepsilon=0.06$ (short dash-dot
line),$\varepsilon=0.1$(solid line) repectively  } \label{fig4}
\end{center}
\end{figure}
\begin{figure}[htb]
\begin{center}
\epsfig{file=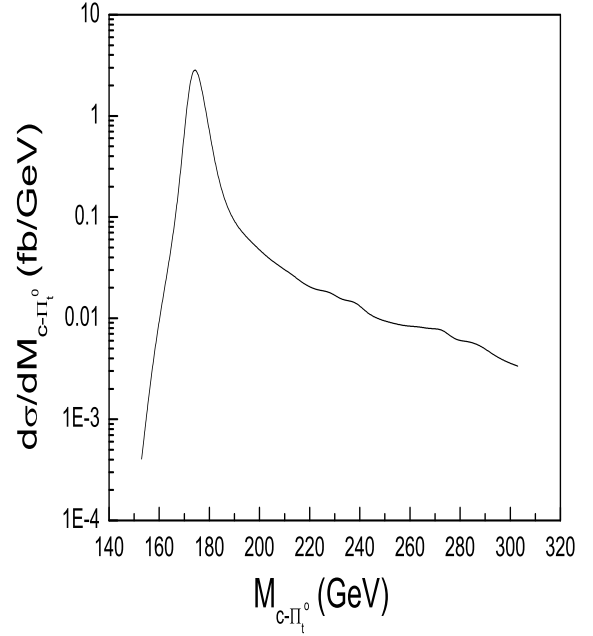,width=240pt,height=280pt} \caption{The
toppion-charm invariant mass distribution for $\sqrt{s}=500$ GeV
 $\varepsilon=0.1$ and $M_{\Pi}=150$ GeV.}
\label{fig5}
\end{center}
\end{figure}
\vspace*{-.2cm}
\null


\end{document}